# Novel Decoding Algorithm for Noiseless Non-Adaptive Group Testing

Manuel Franco-Vivo*,†

**Abstract**

Group testing enables the identification of a small subset of defective items within a larger population by performing tests on pools of items rather than on each item individually. Over the years, it has not only attracted significant attention from the academic community, but has also demonstrated its potential in addressing real-world problems such as infectious disease screening, drug discovery, and manufacturing quality control. With the emergence of the COVID-19 pandemic, interest in group testing has grown further, particularly in non-adaptive testing, due to its time efficiency compared to adaptive approaches. This highlights the importance of improving the performance currently achievable in such a scheme.

This article focuses on advancing the field of noiseless non-adaptive group testing. The main objective of this work is to study and maximize the probability of successfully identifying the subset of defective items while performing as few tests as possible. To this end, we first conduct an in-depth analysis of well-known decoding algorithms used to determine the defectivity status of items, as well as established test design strategies for assigning items to pools. From this review, we identify key opportunities for improvement that inform the development of new decoding algorithms. Specifically, we propose a novel method, Weighted Sequential Combinatorial Orthogonal Matching Pursuit (W-SCOMP), to enhance the efficiency of existing detection procedures. Theoretical results demonstrate that W-SCOMP outperforms other algorithms in noiseless non-adaptive group testing. Furthermore, we develop a simulation framework to model the group testing process and conduct comparative evaluations between the proposed and existing algorithms. The empirical results are consistent with the theoretical findings. Overall, our work expands the range of available decoding algorithms and contributes to the broader understanding of noiseless non-adaptive group testing.

## I. Introduction

GROUP Testing (GT) is concerned with the research of efficient procedures to identify a small subset of items with a condition of interest, referred to as defectives, in a larger population by applying tests on pools of items while minimising the number of tests required.

The first formal proposal of GT is attributed to Robert Dorfman's work on pooled blood screening for syphilis in U.S. Army recruits in 1943 [1], which evinced a substantial drop in the number of assays required, and the related costs, when defectivity is rare. This established a sequential pooled testing paradigm, known as adaptive testing schemes, where later tests are designed based on earlier test outcomes [2]. This differs from non-adaptive testing strategy, in which test pools are fully defined in advance [3]–[5]. For a comprehensive, up-to-date overview of contemporary developments in GT problems, the reader is referred to the monograph of Aldridge, Johnson and Scarlett [3].

Since its inception, GT has evolved mathematically in the broad field of Information Theory, with applications ranging from infectious disease screening [6], [7] and drug discovery [8] to manufacturing quality control [9] and fault diagnosis in networks [10]. More recently, GT has reignited considerable attention with the advent of the ever-expanding bioinformatics technologies, gaining prominence in modern large-scale diagnostic applications [2], [3], [11], such as in the COrona VIrus Disease (Covid-19) pandemic [7], [12]–[17]. In general terms, non-adaptive strategies, which support parallel testing and ensure scalability, can provide significant time savings to high throughput, simpler laboratory automation [18]–[20], presenting therefore potential interest for further research.

This paper focuses on the noiseless non-adaptive group testing model, a foundational framework that assumes error-free test outcomes [3]. In this setting, a test yields a positive result if and only if the test batch contains at least one defective item, and a negative result otherwise. The entire test configuration can be represented as a binary matrix indicating item-to-test assignments. Although this model abstracts away laboratory errors, it renders the problem significantly more mathematically tractable and provides a clear theoretical benchmark that has guided much foundational work (see, e.g. [3], [21]), paving the way for further studies that incorporate the effects of noise [3], [4].

Several studies have demonstrated significant improvements by incorporating novel decoding algorithms developed to accurately determine the defectivity status of items [22]–[25]. Notable non-adaptive models include the Combinatorial Orthogonal Matching Pursuit (COMP) algorithm by Chan et al. [26], as well as the Definite Defectives (DD) and Sequential COMP (SCOMP) algorithms proposed by Aldridge, Johnson, and Baldassini [22].

*Corresponding author.

†School of Mathematics, University of Bristol, UK. Email: pq21381@bristol.ac.uk, mfranco.vivo@gmail.com

Despite these developments, a gap remains between theoretically optimal performance and what is practically achievable with simple, fast decoders under realistic pooling regimes. Motivated by this challenge, the present paper is aimed to address it by improving decoding performance in the noiseless non-adaptive setting while keeping computational complexity appropriate for large-scale applications. Specifically, we propose an enhanced algorithm, Weighted SCOMP (W-SCOMP), that incorporates probabilistic weighting mechanisms to refine the decision-making process during decoding. Theoretical analysis demonstrates that W-SCOMP achieves superior performance compared to existing methods, and simulation results corroborate these findings.

The remainder of the paper is structured as follows. In Section II, we formulate formally the mathematical configuration of the noiseless non-adaptive group testing problem (Subsection II-A) and briefly recall the standard decoding algorithms used (Section II-B). In Section III, we introduce our contributions: the basis behind the novel detection algorithm proposed, detailed procedural description, Weighted SCOMP (Section III-A), present a novel metric named per-test signal-to-noise ratio (SNR) to compare scoring-based group-testing decoders in the noiseless non-adaptive scheme along with its interpretability, limitations and scope (Section III-B), and provide the main theorems of the performance of the Weighted SCOMP (Section III-C), with proofs in appendix. In Section IV-B, we examine the behaviour of the novel algorithm provided by performing simulations. Finally, in Section V, we summarise our concluding remarks.

## II. Notation and Background

Group testing proceeds in two stages: a test *design* which fixes a binary pooling matrix and a *decoder* that estimates the defective set from outcomes [22]. In this work, we adopt the standard noiseless non-adaptive model (see, e.g. [3]). Thus, prior to introducing our contributions, we establish the mathematical framework and the notation we will use, and briefly outline the standard combinatorial decoders used throughout the paper.

### A. Problem Setup

Let $\mathcal{N} = \{1, \ldots, N\}$ be the item set and $\mathcal{K} \subseteq \mathcal{N}$ the unknown set of defective items with $k = |\mathcal{K}|$. Tests are indexed by $\mathcal{T} = \{1, \ldots, T\}$ and the design matrix is denoted by $\mathbf{X} \in \{0,1\}^{T \times N}$ with $X_{t,i} = 1$ iff item $i$ is in test $t$. For each $t \in \mathcal{T}$, we write $T_t = \{j \in \mathcal{N} : X_{t,j} = 1\}$ and $|T_t| = \sum_{j=1}^{N} X_{t,j}$. Under the noiseless model, the outcomes $\mathbf{Y} = (Y_1, \ldots, Y_T) \in \{0,1\}^T$ satisfy

$$Y_t = \mathbb{I}\Big\{ \sum_{j \in T_t} U_j \geq 1 \Big\} = 1 - \prod_{j \in T_t} (1 - U_j), \text{ with } U_j = \mathbb{I}\{j \in \mathcal{K}\},$$

where $\mathbb{I}\{\}$ is the indicator function.

A decoder is a mapping $\widehat{\mathcal{K}} : (\{0,1\}^{T \times N} \times \{0,1\}^T) \to \mathcal{P}(\mathcal{N})$ deployed to infer the unknown defective status of the items. The error probability of completely recovering the defective set $\mathcal{K}$ is given by $\mathbb{P}(\text{err}) = \mathbb{P}\big(\widehat{\mathcal{K}}(\mathbf{X}, \mathbf{Y}) \neq \mathcal{K}\big)$, and under the combinatorial prior, the average error probability is

$$\mathbb{P}(\text{err}_a) = \frac{1}{\binom{N}{k}} \sum_{\mathcal{K}:|\mathcal{K}|=k} \mathbb{P}\big(\widehat{\mathcal{K}}(\mathbf{X}, \mathbf{Y}) \neq \mathcal{K}\big).$$

### B. Decoding algorithms

The standard detection algorithms for group testing used throughout this paper are:

- Combinatorial Orthogonal Matching Pursuit (COMP), which marks as *definite non-defectives* ($\mathcal{DND}$) every item that appears in a negative test. The remaining items form the potential-defective set, $\mathcal{PD}$, which is returned by COMP as $\widehat{\mathcal{K}}_{\text{COMP}} = \mathcal{PD}$. COMP is often used as the initial stage of more sophisticated decoders since it is extremely simple and computationally cheap [26], [27].
- Definite Defectives (DD) algorithm, which refines COMP by using positive-test singletons. After computing $\mathcal{DND}$ and $\mathcal{PD}$ as in COMP, DD labels any item that is the unique member of $\mathcal{PD}$ appearing in a positive test to be a *definite defective* ($\mathcal{DD}$). DD typically improves on COMP, particularly in very sparse regimes [22].
- Sequential COMP (SCOMP), which starts from the DD output and greedily deciphers the remaining set of unexplained positive tests ($\mathcal{T}_u$). At each step, it selects the potential item that covers the most unexplained positive tests, adding it to the estimate defective set $\widehat{\mathcal{K}}_{\text{SCOMP}}$, and updating the set of unexplained positive tests. SCOMP often attains substantially better empirical performance than COMP or DD while remaining computationally feasible [28]–[30].

Detailed algorithmic procedure descriptions can be found in the cited literature.

## III. Main Results

Intuitively, positive tests that comprise fewer items contain more information than positive tests with many participants. Thus, an item appearing in a low-weight positive test has a higher posterior probability of being defective than an item in a high-weight positive test.

Conventional algorithms, such as SCOMP, add candidate defectives to the estimated defective set, by counting the number of unexplained positive tests in which they are present. This binary counting neglects the intrinsic uncertainty of tests that involve multiple possible defectives: a unitary increment treats all unexplained tests equally and can therefore overinflate the score of a candidate that is included in many highly ambiguous tests. To address this issue, we propose an algorithm that incorporates test weights given by the number of potential defectives in a test, $w_t = \sum_{i \in \mathcal{PD}} X_{t,i}$, in order to exploit additional structure in the test matrix and in the distribution of candidate appearances.

### A. Weighted-SCOMP

In the W-SCOMP procedure, each occurrence of a candidate in an unexplained test contributes a fractional value rather than a unit increment. The score of each $i \in \mathcal{PD}$ is obtained as:

$$\mathrm{S}_i^{(\mathcal{T}_u)} = \sum_{t \in \mathcal{T}_u} S_{t,i} = \sum_{t \in \mathcal{T}_u} \frac{1}{w_t^{\alpha}} X_{t,i}.$$

where $\alpha \geq 0$ is a hyperparameter $\alpha$ which is selected after empirical tuning, and controls the influence of the weight on the score. For consistency and comparability, we will fix $\alpha$ as 1 for the remaining of this paper across both theoretical and empirical experiments to allow direct comparison with other decoding algorithms. This normalization gives greater importance to appearances in less ambiguous tests (small $w_t$) while still accounting for the number of unexplained tests in which candidate $i$ appears.

---

**Algorithm 1:** Weighted Sequential Combinatorial Orthogonal Matching Pursuit (W-SCOMP) Algorithm

---

**Input:** Test matrix $\mathbf{X} \in \{0,1\}^{T \times N}$, test outcomes $\mathbf{y} \in \{0,1\}^T$

**Output:** Estimated defective set $\widehat{\mathcal{K}}_{\text{W-SCOMP}}$

**Initialization:**

- Use the DD algorithm to compute the set of definite defectives ($\hat{\mathcal{K}}_{DD}$) and definite non-defectives ($\mathcal{DND}$).
- Initialize the estimated defective set: $\widehat{\mathcal{K}}_{\text{W-SCOMP}} \leftarrow \hat{\mathcal{K}}_{DD}$.
- Set the set of potential defective items: $\mathcal{PD} \leftarrow \{i \in \mathcal{N} : i \notin (\widehat{\mathcal{K}}_{\text{W-SCOMP}} \cup \mathcal{DND})\}$.
- Initialize the set of unresolved tests: $\mathcal{T}_u \leftarrow \{t \in \mathcal{T} : y_t = 1 \text{ and}, \forall i \in t, i \notin \widehat{\mathcal{K}}_{\text{W-SCOMP}}\}$.

**Iterative Defective Identification:**

- **Step 1:** For each test $t \in \mathcal{T}_u$, compute the inverse of its row weight:

$$w_t \leftarrow \sum_{i \in \mathcal{PD}} X_{t,i}.$$

- **Step 2:** Compute the score for each item, and select the item $\hat{i} \in \mathcal{PD}$ with the maximum score:

$$S_i = \sum_{t \in \mathcal{T}_u} X_{t,i} * \frac{1}{w_t}$$

  Identify the item $\hat{i} \in \mathcal{PD}$ that best explains the unresolved positive tests:

$$\hat{i} = \arg\max_{i \in \mathcal{PD}} S_i.$$

- **Step 3:** Add $\hat{i}$ to the estimated defective set:

$$\widehat{\mathcal{K}}_{\text{W-SCOMP}} \leftarrow \widehat{\mathcal{K}}_{\text{W-SCOMP}} \cup \{\hat{i}\}.$$

- **Step 4:** Update the set of potential defectives and unresolved tests:

$$\mathcal{T}_u \leftarrow \{t \in \mathcal{T}_u : y_t = 1 \text{ and}, \forall i \in t, i \notin \widehat{\mathcal{K}}_{\text{W-SCOMP}}\}.$$

$$\mathcal{PD} \leftarrow \{i \in \mathcal{PD} : \exists t \in \mathcal{T}_u \text{ where } X_{t,i} = 1\}.$$

**Termination:** Repeat the iterative process until either:

- $\mathcal{T}_u = \emptyset$, or
- No further items can be identified as defectives.

**return** $\widehat{\mathcal{K}}_{\text{W-SCOMP}}$

---

### B. Signal-to-Noise Ratio as a Metric for Comparing Scoring-based Decoders

We propose a signal-to-noise ratio (SNR) as a metric to compare scoring-based group-testing decoders, for instance, the refinement stage of Weighted SCOMP vs. traditional SCOMP. Henceforth, we consider a fixed item $i$. For a given scoring decoder, denoting by $S_{t,i} \in \mathbb{R}$ the per-test contribution assigned to the item $i$ in test $t \in \mathcal{T}$, we test

$$H_D : i \text{ is defective } (i \in D) \quad \text{vs.} \quad H_{ND} : i \text{ is non-defective } (i \in ND),$$

based on the aggregated score

$$S_i^{(\mathcal{T})} = \sum_{t \in \mathcal{T}} S_{t,i}.$$

For this purpose, we adopt two standard assumptions for scoring-based decoders.

**Assumption 1** (Additive local scores)**.** *The decision statistic for item $i$ is a sum of per-test contributions,*

$$S_i^{(\mathcal{T})} = \sum_{t \in \mathcal{T}} s\big(X_{t,i}, Y_t\big)$$

*where $s : \{0,1\} \times \{0,1\} \to \mathbb{R}$ depends only on the single-test outcome $(X_{t,i}, Y_t)$ and is upper-bounded.*

**Assumption 2** (Finite second moments and variance additivity)**.** *Given a condition $c$, the per-test random variables $s\big(X_{t,i}, Y_t\big)$ with $t \in \mathcal{T}$, each having finite conditional mean and variance, $\mathbb{E}[s\big(X_{t,i}, Y_t\big) \mid c] < \infty$, $\mathbb{E}[s\big(X_{t,i}, Y_t\big)^2 \mid c] < \infty$, satisfy*

$$\mathrm{Var}\left[\sum_{t \in \mathcal{T}} s\big(X_{t,i}, Y_t\big) \mid c\right] = \sum_{t \in \mathcal{T}} \mathrm{Var}\big[s\big(X_{t,i}, Y_t\big) \mid c\big].$$

Under Assumptions 1 and 2 and the independence between tests, per-test class means and variances can be expressed as

$$\mu_c := \mathbb{E}[S_{t,i} \mid c], \qquad \sigma_c^2 := \mathrm{Var}[S_{t,i} \mid c], \quad c \in \{D, ND\}.$$

From the per-test mean gap $\Delta\mu := \mu_D - \mu_{ND}$, and the combined per-test variance $\Sigma^2 := \sigma_D^2 + \sigma_{ND}^2$, we then define the *Signal-to-Noise Ratio* by

$$\mathrm{SNR}_{\mathcal{T}} := \frac{\mathbb{E}[S_i^{(\mathcal{T})} \mid D] - \mathbb{E}[S_i^{(\mathcal{T})} \mid ND]}{\sqrt{\mathrm{Var}[S_i^{(\mathcal{T})} \mid D] + \mathrm{Var}[S_i^{(\mathcal{T})} \mid ND]}},$$

which can be rewritten as

$$\mathrm{SNR}_{\mathcal{T}} = \sqrt{T}\,\mathrm{SNR}_{\mathrm{per}},$$

where $\mathrm{SNR}_{\mathrm{per}} = \frac{\Delta\mu}{\sqrt{\Sigma^2}}$ is named the per-test signal-to-noise ratio.

We now give the theorems with outlines of their proofs, which will allow us to interpret the SNR. The full proofs are deferred to Appendix A.

**Theorem 1.** *Under Assumptions 1 and 2, the midpoint threshold classifier $\theta = \frac{1}{2}T(\mu_D + \mu_{ND})$ satisfies*

$$\mathbb{P}(\mathrm{err}_{\mathcal{T}}) \le \frac{4}{\mathrm{SNR}_{\mathcal{T}}^2}.$$

*Proof.* Apply Chebyshev's inequality to bound each tail probability (false negative and false positive) using $\mathrm{Var}[S_i^{(T)} \mid c] = T\sigma_c^2$, and we sum the two bounds for $c \in \{D, ND\}$. More details are provided in Appendix A1. □

**Theorem 2.** *Under Assumptions 1 and 2, the Bayes error, $\mathrm{err}_{\mathcal{T}}^*$, obeys*

$$\mathrm{err}_{\mathcal{T}}^* < \tfrac{1}{2}\exp\Big(-\frac{\mathrm{SNR}_{\mathcal{T}}^2}{4}\Big).$$

*Proof.* Evaluate Bhattacharyya's distance $B$ between the two Gaussian approximations of $(S_i^{(\mathcal{T})}|c)$ for $c \in \{D, ND\}$ and apply the standard bound $\mathrm{err}_{\mathcal{T}}^* \le \frac{1}{2}e^{-B}$. Details can be found in Appendix A2. □

Theorems 1–2 imply that, for additive finite-variance scoring decoders, larger SNR yields improved upper bound for probability of error and Bayes error, i.e., better decoding algorithm.

Nevertheless, some decoding algorithms such as DD are not generally comparable via per-test SNR. This decoder's decisions are inherently combinatorial and non-local, and therefore violate Assumption 1.

**Proposition 1.** *DD's decision for an item depends on the global candidate set and thus cannot be represented as a function solely of per-test $(X_{t,i}, Y_t)$ contributions.*

*Proof.* The singleton-test condition used by DD depends on whether other items are candidates. This is a global property and cannot be encoded by local per-test functions of $(X_{t,i}, Y_t)$. A formal construction is given in Appendix A3. □

In summary, the per-test SNR is a rigorous and practical metric for scoring-based decoders under additivity and finite-variance assumptions, since it quantifies the extent to which the average score of defective items is separated from that of non-defective items relative to the variability of the scores. A higher SNR indicates that the two distributions are well separated, i.e., the "signal" (difference in means) dominates the "noise" (standard deviation) leading to more reliable discrimination between defectives and non-defectives.

This metric is particularly useful in the theoretical evaluation of group testing algorithms for the following reasons:

- **Quantitative Measure of Separability:** The SNR directly measures how much the defectives' scores exceed those of non-defectives relative to the fluctuations in both scores. Hence, a larger SNR implies a lower probability of misclassification.
- **Incorporation of Variance:** By accounting for the variance in scores, the SNR provides a more robust performance metric than a simple difference in means. Even if the difference in means is modest, a low variance (small noise) can lead to a high SNR, indicating reliable detection.
- **Analytical Comparability:** The explicit expressions for the first and second moments of the scores enable us to derive closed-form (or tractable) expressions for the SNR for different rules. This facilitates comparisons under the same testing conditions.

However, it is not universally applicable to combinatorial decoders such as DD without additional approximation arguments.

### C. Theoretical Performance

In order to theoretically establish the performance improvement of the novel Weighted SCOMP (W-SCOMP) algorithm over other well-known decoding procedures, our analysis concentrates on the fundamental task of discriminating between defective and non-defective items. Since SCOMP has already been shown, both theoretically and empirically, to perform at least as well as other optimal decoding algorithms (see, e.g., [3]), a direct comparison between W-SCOMP and SCOMP is particularly instructive. By analyzing the extent to which each algorithm separates the score distributions of defective and non-defective items, we obtain a precise measure of their detection capabilities. In this context, the SNR serves as a natural and rigorous metric for quantifying this separation and the likelihood of misclassfication. Consequently, by demonstrating that W-SCOMP attains a higher SNR than SCOMP, we provide strong theoretical evidence of its enhanced performance.

For this purpose, we consider a Bernoulli test design matrix $\mathbf{X}$ ($X_{t,i} \sim \text{Bernoulli}(p)$) so that each item $i$ appears in each test independently with probability $p$. The choice of $p$ is designed to balance sparsity and information: as the number of unknown defective items $k$ increases, each item is included in fewer tests, thereby keeping tests sparse and reducing overlaps among defectives. We use the theoretically optimal $p = \frac{1}{k+1}$, which minimises the upper bound of the error probability given in Eq. (2.9) of [3]. The Bernoulli design is widely used in the group testing literature, see [24], [27], [31]–[35] among others.

From the notation given in Section II, when focusing on a specific item $i$, we write $Z_t^{(i)} := \sum_{\substack{j \in \mathcal{PD} \\ j \neq i}}^{N} X_{t,j}$, so that if item $i$ is included in $T_t$ then $w_t = 1 + Z_t^{(i)}$.

*1) Moments for the Score of Weighted and Unweighted SCOMP:*

*a) Weighted SCOMP:* The contribution $S_{t,i}$ for item $i$ in test $t$ assigned by the Weighted SCOMP rule is denoted as

$$W_{t,i} \;=\; \frac{1}{w_t}\,\mathbb{I}\{X_{t,i} = 1 \text{ and } Y_t = 1\}, \tag{1}$$

where $Y_t = 1$ denotes that test $t$ is positive (i.e., at least one defective is present) and $w_t$ is the number of potential defective items included in test $t$.

We treat separately the cases where item $i$ is defective ($i \in D$) and non-defective ($i \in ND$). The following proposition summarizes the per-test weighted mean and second moment.

**Proposition 2.** *Consider $N$ items with $k$ defectives. Let each item be independently included in a test with probability $p$. For a single test $t$ and item $i \in D$, then*

$$\mu_D^{(w)}(N,k) = \mathbb{E}\big[W_{t,i} \,\big|\, i \in D\big] = \frac{1-(1-p)^N}{N}, \tag{2}$$

$$\nu_D^{(w)}(N,k) = \mathbb{E}\big[W_{t,i}^2 \,\big|\, i \in D\big] = p\sum_{j=0}^{N-1} \frac{1}{(1+j)^2}\binom{N-1}{j} p^j (1-p)^{N-1-j}, \tag{3}$$

*and, when $i$ is non-defective, we have*

$$\mu_{ND}^{(w)}(N,k) = \mathbb{E}\big[W_{t,i} \,\big|\, i \in ND\big] = p\,q(k)\,\mu_{ND}(N,k), \tag{4}$$

$$\nu_{ND}^{(w)}(N,k) = \mathbb{E}\big[W_{t,i}^2 \,\big|\, i \in ND\big] = p\,q(k)\,\nu_{ND}(N,k), \tag{5}$$

*where $q(k) = 1-(1-p)^k$, and $\mu_{ND}(N,k)$ and $\nu_{ND}(N,k)$ are given by (21) and (22) respectively, in Appendix B.*

*Proof.* For a defective item $i$, whenever $X_{t,i} = 1$ the test is necessarily positive, hence (1) can be rewritten as $W_{t,i}^{(D)} = \frac{1}{1+Z_t}\mathbb{I}\{X_{t,i} = 1\}$ where $Z_t \sim \mathrm{Bin}(N-1, p)$ counts the other items in the test. Taking expectations and using the binomial identity yields (2) and (3), see details in Appendix B1.

For a non-defective item $i$, let $Z_{t,D} \sim \mathrm{Bin}(k, p)$ be the number of defectives (among the other items) in test $t$ and $Z_{t,ND} \sim \mathrm{Bin}(N-k-1, p)$ be the number of other non-defectives, which are independent. Conditioning on the event $A_t = \{Z_{t,D} \geq 1\}$ (the test is positive) and expanding the conditional expectation, the expressions (4)–(5) are obtained. The detailed expansions are shown in Appendix B2. □

**Proposition 3.** *For a single test $t$ and an item $i$, let $\mu_D^{(w)}(N,k), \mu_{ND}^{(w)}(N,k)$ and $\nu_D^{(w)}(N,k), \nu_{ND}^{(w)}(N,k)$ be the per-test first and second moments for defective and non-defective items defined in Proposition 2. Then:*

$$\begin{aligned}\Delta\mu^{(w)}(N,k) &= \mu_D^{(w)}(N,k) - \mu_{ND}^{(w)}(N,k) = p\big[\mu_D(N,k) - q(k)\,\mu_{ND}(N,k)\big], && (6)\\ \sigma_w^2(N,k) &= \big(\nu_D^{(w)} - \mu_D^{(w)2}\big) + \big(\nu_{ND}^{(w)} - \mu_{ND}^{(w)2}\big)\\ &= p\Big[\big(\nu_D(N,k) - p\mu_D^2(N,k)\big) + q(k)\big(\nu_{ND}(N,k) - pq(k)\mu_{ND}^2(N,k)\big)\Big], && (7)\end{aligned}$$

*and*

$$\mathrm{SNR}_W(N,k) = \frac{\Delta\mu^{(w)}(N,k)}{\sqrt{\sigma_w^2(N,k)}}. \tag{8}$$

*b) Unweighted SCOMP:* In the unweighted SCOMP, the contribution of item $i$ in test $t$ is the indicator that the item appears in a positive test, denoted by

$$U_{t,i} = \mathbb{I}\{X_{t,i} = 1 \text{ and } Y_t = 1\}. \tag{9}$$

As in the weighted case, we treat defective and non-defective items separately. The following proposition summarizes the per-test first and second moments.

**Proposition 4.** *Under independent test inclusion with probability $p$. For a single test $t$ and item $i$,*

$$\begin{aligned}\mu_D^{(u)}(k) &= \mathbb{E}\big[U_{t,i}\,\big|\, i \in D\big] = p, && (10)\\ \nu_D^{(u)}(k) &= \mathbb{E}\big[U_{t,i}^2\,\big|\, i \in D\big] = p, && (11)\\ \mu_{ND}^{(u)}(k) &= \mathbb{E}\big[U_{t,i}\,\big|\, i \in ND\big] = p\,q(k), && (12)\\ \nu_{ND}^{(u)}(k) &= \mathbb{E}\big[U_{t,i}^2\,\big|\, i \in ND\big] = p\,q(k), && (13)\end{aligned}$$

*where $q(k) = 1 - (1-p)^k$ is the probability that a test contains at least one defective conditioned on any given non-defective item.*

*Proof.* For a defective item, the inclusion $X_{t,i} = 1$ implies that the test is positive, hence (9) can be rewritten as $U_{t,i}^{(D)} = \mathbb{I}\{X_{t,i} = 1\}$, and the first and second moments equal $\mathbb{P}(X_{t,i} = 1) = p$. For a non-defective item, positivity requires that at least one defective among the remaining $k$ items is included. Therefore, $\mathbb{P}(X_{t,i} = 1 \text{ and } Y_t = 1) = p\,q(k)$ and, since $U_{t,i}$ is an indicator variable, the second moment equals the first. See further, in Appendix C. □

**Proposition 5.** *From the moments given in Proposition 4, we obtain*

$$\begin{aligned}\Delta\mu^{(u)}(k) &= \mu_D^{(u)}(k) - \mu_{ND}^{(u)}(k) = p\big(1 - q(k)\big), && (14)\\ \sigma_u^2(k) &= \big(\nu_D^{(u)} - \mu_D^{(u)2}\big) + \big(\nu_{ND}^{(u)} - \mu_{ND}^{(u)2}\big) = p(1-p) + p\,q(k)\big(1 - p\,q(k)\big), && (15)\end{aligned}$$

*and*

$$\mathrm{SNR}_U(k) = \frac{\Delta\mu^{(u)}(k)}{\sqrt{\sigma_u^2(k)}} = \sqrt{p}\,\frac{1 - q(k)}{\sqrt{1 - p + q(k)\big(1 - p\,q(k)\big)}}. \tag{16}$$

*2) SNR comparison:*

**Theorem 3.** *The per-test signal-to-noise ratio for the weighted rule is greater or equal to the one for the unweighted rule, i.e.,*

$$SNR_W(N,k) \geq SNR_U(k)$$

*for all integer $k \geq 1$ and $N > k$.*

*Proof.* We consider the normalized ratios $R_W(N,k)$ and $R_U(k)$ defined in (23)–(24), since $\mathrm{SNR}_W(N,k) = (p/\sqrt{p})\,R_W(N,k)$ and $\mathrm{SNR}_U(k) = (p/\sqrt{p})\,R_U(k)$, and thus $\mathrm{SNR}_W(N,k) \geq \mathrm{SNR}_U(k)$ is equivalent to

$$R_W(N,k) \geq R_U(k). \tag{17}$$

For this purpose, we introduce the simplified random-variable notation used throughout the derivation: $W_D = (1 + Z_t)^{-1}$ and $W_{ND} = (1 + Z_{t,D} + Z_{t,ND})^{-1} \mid Z_{t,D} \geq 1)$ independent. The relevant first and second moments appear in Proposition 2 and are collected in Appendix D1 for convenience.

The proof proceeds by the following steps.

(i) Compute the numerator $\mathbb{E}[W_D] - q(k)\mathbb{E}[W_{ND}]$ of (23) explicitly. This calculation reduces to the closed formula (29) in Appendix D2, which shows the numerator is strictly positive for every $0 < k < N$.

(ii) Bound the second moments using Jensen's inequality to obtain the lower bounds displayed in Appendix D4. These bounds allow control over the denominator of $R_W(N, k)$ given in (23).

(iii) Square both sides of inequality (17) (both sides are non-negative) and rearrange to obtain an equivalent inequality of the form

$$\Big(\frac{\mathbb{E}[W_D] - q(k)\mathbb{E}[W_{ND}]}{1 - q(k)}\Big)^2 \big(1 - p + q(k)(1 - pq(k))\big) \geq \mathbb{E}[W_D^2] - p\,\mathbb{E}[W_D]^2 + q(k)\big(\mathbb{E}[W_{ND}^2] - pq(k)\mathbb{E}[W_{ND}]^2\big). \quad (18)$$

Expanding and collecting terms, one can reach the inequality

$$0 \leq \mathbb{E}[W_D]^2 f_1(k) + \mathbb{E}[W_D]\mathbb{E}[W_{ND}]f_2(k) + \mathbb{E}[W_{ND}]^2 f_3(k) + \big(\mathbb{E}[W_D^2] + q(k)\mathbb{E}[W_{ND}^2]\big)f_4(k), \quad (19)$$

where $f_1(k), \ldots, f_4(k)$ are the functions given in Appendix D6.

(iv) Substitute the explicit first-moment expressions $\mu_D^2(N, k)$ and $\mu_{ND}^2(N, k)$ given in (20) and (21) and the exact representation

$$\mathbb{E}[W_D^2] + q(k)\mathbb{E}[W_{ND}^2] = \frac{2}{N}\sum_{s=1}^{N}\binom{N}{s}\frac{1}{s}p^{s-1}(1-p)^{N-s} - \frac{1}{N-k}\sum_{s=1}^{N-k}\frac{1}{s}\binom{N-k}{s}p^{s-1}(1-p)^{N-s}$$

(see Appendix D5) into (19) and define the resulting function as $f(N, k)$, which is displayed in Appendix D6.

(v) For every fixed $k \geq 1$, $f_1(k), \ldots, f_4(k)$ do not depend on $N$, and one can check that $\lim_{N\to\infty} f(N, k) = 0$, see in Appendix D7. Hence, the inequality (19) holds iff $f(N, k) \geq 0$, this is empirically checked and illustrated by Figures 10 and 11 for a wide range of $(N, k)$ values, hence (19) holds for all admissible $(N, k)$.

Combining (i)–(v) proves (17), and therefore $\mathrm{SNR}_W(N, k) \geq \mathrm{SNR}_U(k)$ for all integers $k \geq 1$ and $N > k$.

□

This theorem guarantees that the Weighted SCOMP rule never underperforms, and typically outperforms, its unweighted counterpart (SCOMP) in per-test SNR. In practice, this implies fewer tests are required to achieve the same detection accuracy when using the weighted contributions.

**Corollary 1.** *Let $W_{t,i}$ and $U_{t,i}$ be the per-test weighted and unweighted contributions given by (1) and (9). Let $W_i^{(\mathcal{T})}$ and $U_i^{(\mathcal{T})}$ be the aggregated scores for an item $i$ over $T$ independent tests defined by*

$$W_i^{(\mathcal{T})} = \sum_{t\in\mathcal{T}} W_{t,i}, \quad U_i^{(\mathcal{T})} = \sum_{t\in\mathcal{T}} U_{t,i},$$

*Let $\mu^{(w)}(N, k)$ and $\mu^{(u)}(k)$ denote the per-test means and $\sigma_w^2(N, k)$ and $\sigma_u^2(k)$ the per-test variances of $W_{t,i}$ and $U_{t,i}$, respectively. If there exist bounds $M_w$ and $M_u$ such that $M_w \leq M_u$ and almost surely*

$$|W_{t,i} - \mu^{(w)}(N, k)| \leq M_w \quad \text{and} \quad |U_{t,i} - \mu^{(u)}(k)| \leq M_u,$$

*then the variance $\sigma_w^2(N, k)$ is smaller (or at least not larger) than $\sigma_u^2(k)$.*

Corollary 1 can be interpreted as the weighted score $W_i^{(\mathcal{T})}$ exhibits a sharper concentration around its mean compared to $U_i^{(\mathcal{T})}$ see proof in Appendix E. From a practical point of view, for a given number of tests $T$, the probability that the Weighted SCOMP score deviates from its expected value by a fixed amount $\varepsilon$ decays exponentially faster than for the unweighted SCOMP score. This improved concentration implies a lower probability of misclassifying an item when using the W-SCOMP rule.

## IV. Performance Evaluation

This section empirically evaluates the decoding algorithms discussed earlier. A Python-based simulation framework was developed to test their performance under varying test matrix constructions and defect sparsity levels.

### A. Simulation

A critical aspect of this analysis is the statistical model assumed for the defective set $\mathcal{K}$. There are two common models used in prior works on noiseless non-adaptive group testing: *Combinatorial Prior*, where number of defective items is fixed at $k$, and the defective set is uniformly distributed over all subsets of $\{1, 2, \ldots, N\}$ of size $k$; and *Independent and Identically Distributed Prior*, where each item is defective independently with a fixed probability $p$ [3].

Since combinatorial prior supports worst-case analysis by providing uniform performance guarantees across all defective sets of size $k$ [21], and we aim to study performance for fixed values of $k$, the combinatorial prior is adopted. It provides a clear structure for deriving performance bounds and algorithmic design. Although $k$ is used to construct $\mathcal{K}$, the decoding algorithms operate without prior knowledge or estimation of this value.

### B. Empirical Results

This section compares the performance of the proposed and existing decoding algorithms using the simulation results and derived metrics.

*1) Success Probability for Exact Recovery:* Figures 1–6 include the *counting bound* proposed by Baldassini, Johnson and Aldridge in [36], which states that for any decoding algorithm performing $T$ tests, the probability of success for recovering the defective set with $T$ tests, in any scenario, satisfies

$$\mathbb{P}(suc) \leq \frac{2^T}{\binom{N}{k}}$$

This is presented as a dashed line, representing the information-theoretic upper limit on success probability. This enables a direct comparison between the empirical success probabilities of the algorithms and this theoretical bound.

*2) Exact Recovery Criterion:* A decoding algorithm estimates the defective set $\hat{\mathcal{K}}$ from $\mathbf{Y}$ and $\mathbf{X}$. *Exact recovery* is achieved when $\hat{\mathcal{K}} = \mathcal{K}$. The empirical success probability is defined as

$$\mathbb{P}(suc) = \frac{1}{n} \sum_{j=1}^{n} \mathbb{I}\{\hat{\mathcal{K}}_j = \mathcal{K}_j\},$$

where $\{(\hat{\mathcal{K}}_j, \mathcal{K}_j)\}_{j=1}^{n}$ denote the estimated and true defective sets over $n$ independent trials. This rigorous criterion allows exploration of the fundamental limits of group testing and offers insight into the minimum number of tests $T$ required to recover $\mathcal{K}$ with high probability.

Figure 1 compares five decoding algorithms under a Bernoulli design with $N = 500$ and $k = 10$, using the optimal probability $p = \frac{1}{k+1}$. Empirical results show that W-SCOMP slightly outperforms the others in exact recovery, followed closely by SCOMP, while DD and COMP perform comparably but with lower accuracy. Moreover, Figure 1 also provides a detailed view for $75 \leq T \leq 125$, illustrating the subtle advantage of W-SCOMP.

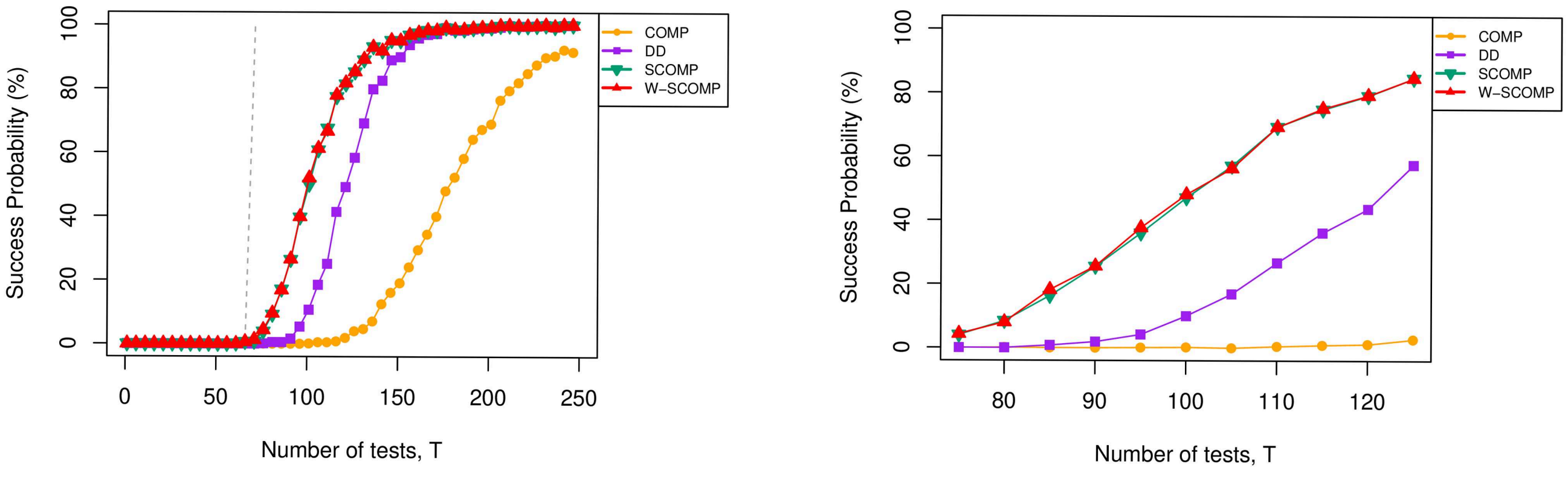

Fig. 1. Average success probability over $n = 1000$ independent simulations for each $T$ in a Bernoulli test design with $N = 500$ and $k = 10$ (left), and zoomed in version for $T \in [75, 125]$ (right).

*3) Different Test Designs:* The decoding algorithms are further compared across different design configurations to quantitatively evaluate their performance trade-offs and robustness under diverse channel and system conditions. The test designs used are the Constant Tests per Item Design where each item participates in exactly $L$ tests chosen uniformly at random without replacement, ensuring that columns are independent and every column of $\mathbf{X}$ has exactly $L$ ones, $\sum_{t=1}^{T} X_{t,i} = L$ for all $i \in \mathcal{N}$, as studied in [23], [29]; and the Near-Constant Tests per Item Design, where $L$ tests are chosen uniformly at random with replacement for each item $i$ [18]. This makes the design more mathematically tractable compared to constant tests per item design [18], [24], and has received ample attention in the literature, see among others [24], [37], [38]. For both designs we use $L = \left\lfloor \frac{T}{k} \ln 2 \right\rfloor$ as the optimal choice that maximises information per test, see Theorem 2 in [24].

In both Constant (Figure 2) and Near-Constant (Figure 3) Tests per Item designs, all algorithms achieve higher success probabilities than under the Bernoulli design. As expected from their construction, the performance difference between these two designs is minimal. Moreover, the enlarged versions in Figures 2 and 3 show that W-SCOMP attains slightly higher success probability than SCOMP, followed by DD and COMP, mirroring the trend observed in the Bernoulli design.

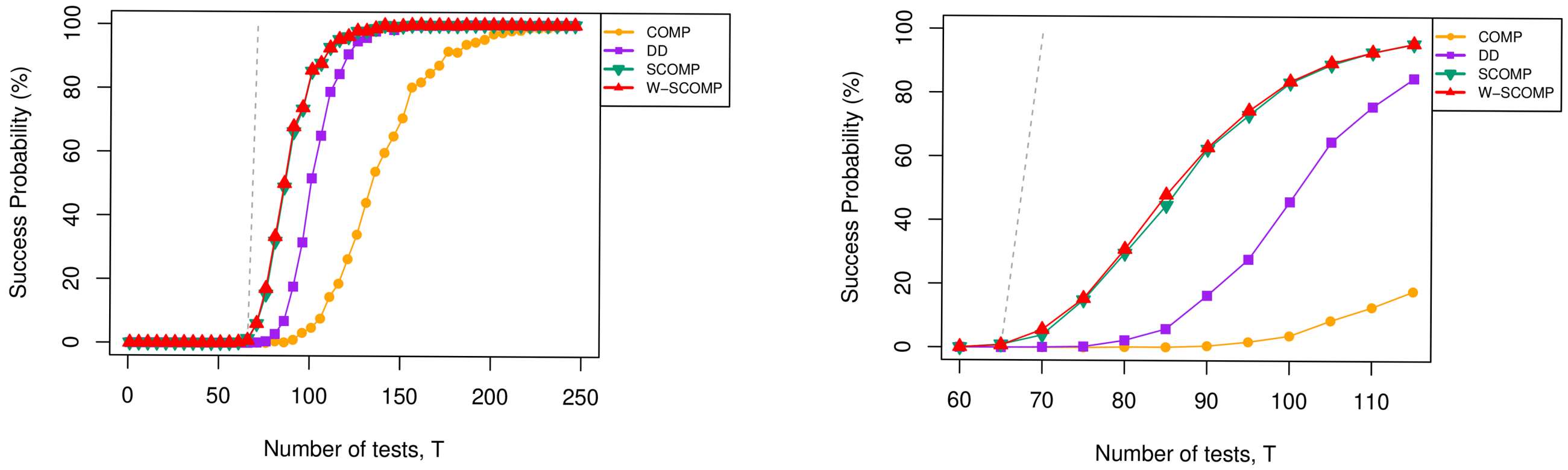

Fig. 2. Average success probability over $n = 1000$ independent simulations for each $T$ in a Constant Tests per Item design with $N = 500$ and $k = 10$ (left), and zoomed in version for $T \in [60, 115]$ (right).

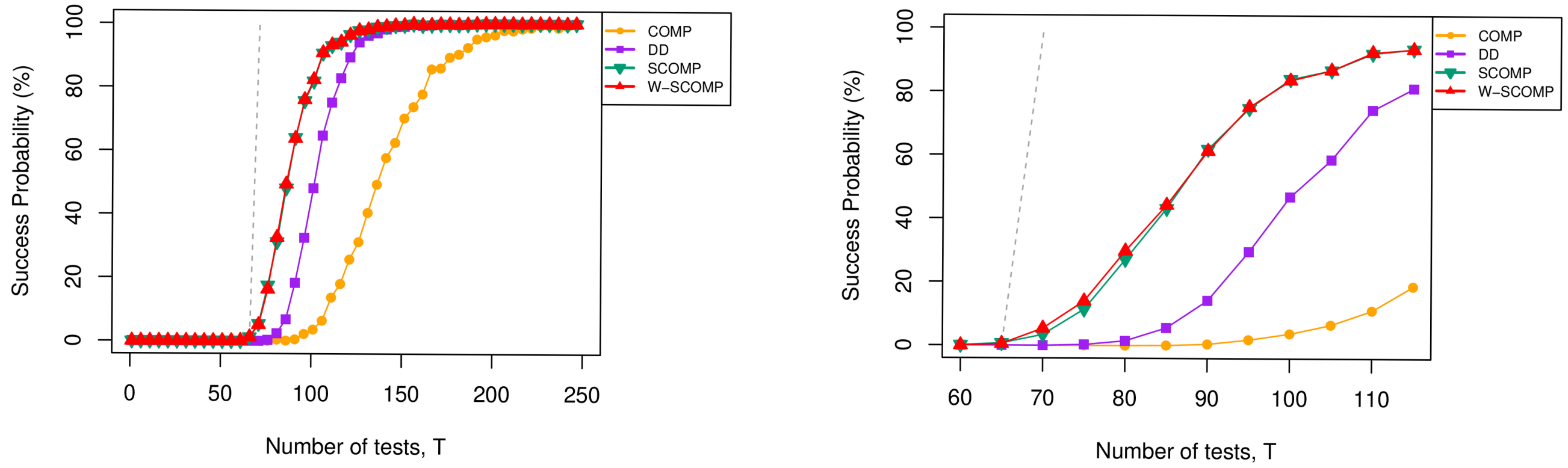

Fig. 3. Average success probability over $n = 1000$ independent simulations for each $T$ in a Near-Constant Tests per Item design with $N = 500$ and $k = 10$ (right), and zoomed in version for $T \in [60, 115]$ (right).

*4) Varying Defect Density:* We next investigate how varying defective sparsity affects recovery performance. The left panels in Figures 4, 5, and 6 show results for a sparser setting ($k = 5$), while the right panels correspond to a denser regime ($k = 25$). Overall, sparser designs yield more efficient recovery, consistent with the counting bound, which predicts that smaller $k$ reduces the number of tests required for exact recovery. In terms of algorithm performance, we observe the same pattern as in previous experiments, where W-SCOMP slightly outperforms SCOMP followed by DD and COMP.

*5) Error Analysis of Recovered Set:* A key component of evaluating decoding algorithms for noiseless non-adaptive group testing is the analysis of errors in the recovered set, in particular false negatives and false positives. These error types have different operational implications and their reduction is essential for practical deployments.

A false negative occurs when a defective item is omitted from the estimate, i.e., any element of $\mathcal{K} \setminus \widehat{\mathcal{K}}$. Such errors can be critical in safety or health-sensitive applications. The left plot of Figure 7 presents the average number of false negatives produced by each decoding algorithm under the Bernoulli design. By construction, COMP's estimate always contains the true defective set, hence its false-negative curve is identically zero [22]. When the number of tests is small, the remaining algorithms frequently lack sufficient information and therefore miss many defectives. W-SCOMP and SCOMP reduce false negatives more rapidly than DD, indicating more efficient use of available tests. As $T$ increases, all algorithms that are initialised with DD converge to zero false negatives, since DD recovers an increasing portion of the true defectives and the subsequent refinement steps become less influential.

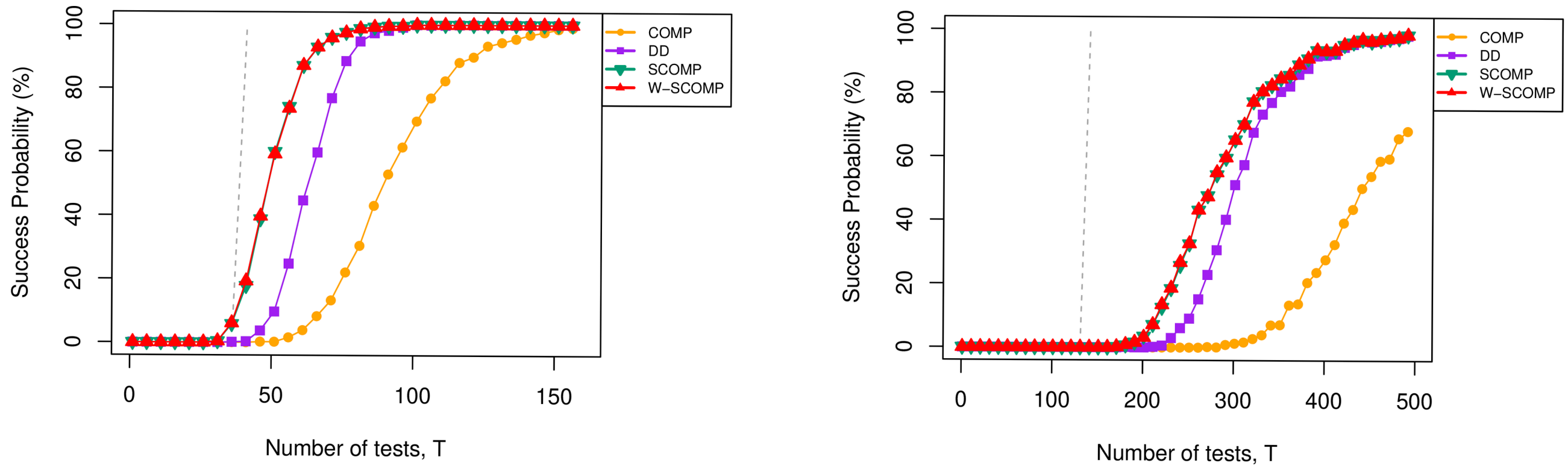

Fig. 4. Average success probability over $n = 1000$ independent simulations for each $T$ in a Bernoulli test design with $N = 500$ and $k = 5$ (left), and $k = 25$ (right).

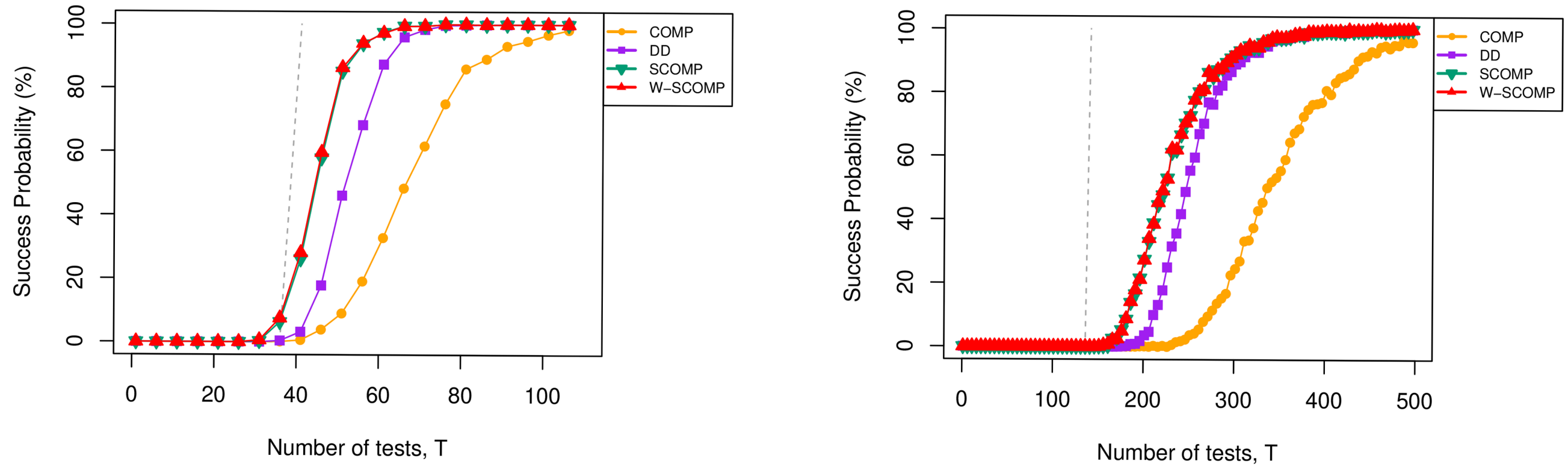

Fig. 5. Average success probability over $n = 1000$ independent simulations for each $T$ in a Constant Tests per Item design with $N = 500$ and $k = 5$ (left), and $k = 25$ (right).

Conversely, a false positive occurs when a non-defective item is declared defective, i.e., an element of $\hat{\mathcal{K}} \setminus \mathcal{K}$. Although false positives generally pose smaller safety risks than false negatives, they increase follow-up testing and operational cost, which can impair large-scale screening efficiency.

The right hand side of Figure 7 reports average false positives across algorithms. DD never produces false positives because its output is always a subset of the true defective set [22]. COMP, which returns the set of possible defectives, exhibits substantially higher false-positive counts. W-SCOMP and SCOMP initially include additional candidates to avoid missing defectives, which raises false positives at moderate $T$. However, as $T$ grows, these algorithms progressively eliminate spurious items and their false-positive rates decline to zero. Overall, W-SCOMP yields a favorable trade-off between false negatives and false positives, supporting the theoretical comparison in Theorem 3 that W-SCOMP performs at least as well as SCOMP in distinguishing defectives from non-defectives.

### C. *Comparison between $\hat{\mathcal{K}}$ and $\mathcal{K}$*

To quantify how closely an estimated set $\hat{\mathcal{K}}$ approximates the true set $\mathcal{K}$, we employ two standard set-similarity metrics: the *Jaccard index* [39] and the *F-score* [40].

The Jaccard index is defined as

$$J(\mathcal{K}, \widehat{\mathcal{K}}) = \frac{|\mathcal{K} \cap \widehat{\mathcal{K}}|}{|\mathcal{K} \cup \widehat{\mathcal{K}}|},$$

this index measures overlap relative to the union and penalizes both false positives and false negatives, making it well suited for sparse defective sets.

The $F_1$-score is the harmonic mean of the precision and recall indexes

$$F_1 = 2 \cdot \frac{Precision \cdot Recall}{Precision + Recall},$$

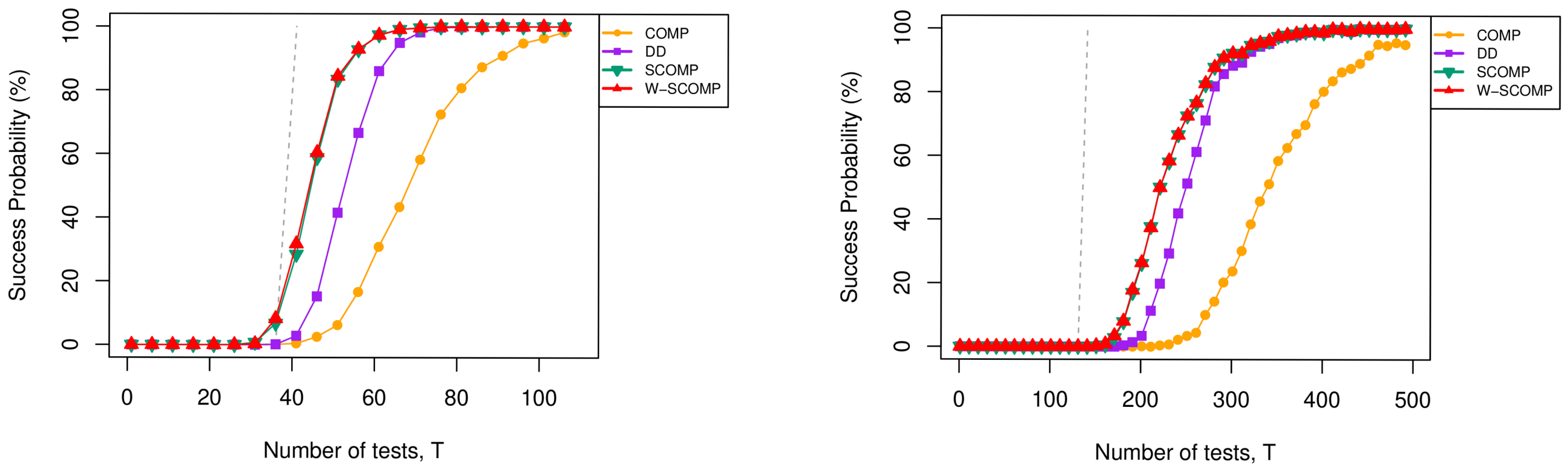

Fig. 6. Average success probability over $n = 1000$ independent simulations for each $T$ in a Near-Constant Tests per Item design with $N = 500$ and $k = 5$ (left), and $k = 25$ (right).

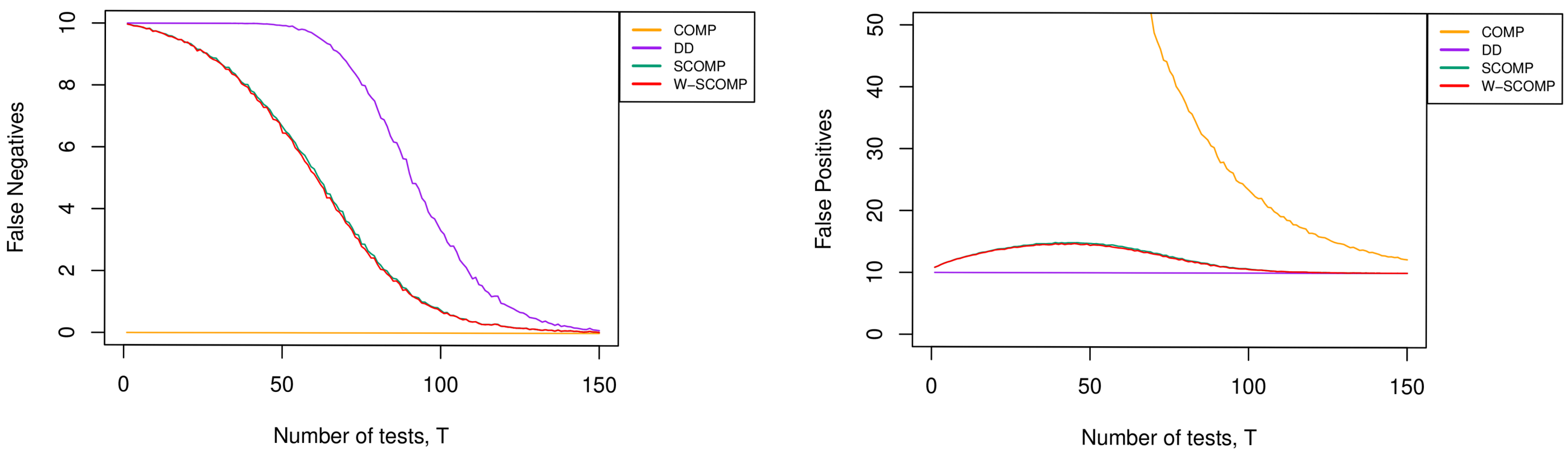

Fig. 7. Average errors in recovered set $\widehat{\mathcal{K}}$ for $T$ tests over $n = 1000$ independent simulations for each $T$ in a Bernoulli test design with $N = 500$ and $k = 10$: False Negatives (left) and False Positives (right).

where

$$\mathit{Precision} = \frac{|\mathcal{K} \cap \widehat{\mathcal{K}}|}{|\widehat{\mathcal{K}}|} \quad \text{and} \quad \mathit{Recall} = \frac{|\mathcal{K} \cap \widehat{\mathcal{K}}|}{|\mathcal{K}|},$$

the $F_1$-score balances the trade-off between false positives and false negatives, and is therefore appropriate when both error types are important.

Both metrics shown in Figure 8 exhibit similar trends. SCOMP and W-SCOMP consistently approximate $\mathcal{K}$ more accurately than DD and COMP, with W-SCOMP showing a slight advantage. For small $T$, DD's near-empty estimates yield lower Jaccard and $F_1$-score, reflecting its conservative behaviour.

### *D. W-SCOMP and SCOMP Comparison*

A natural form in which to compare W-SCOMP and SCOMP is by the *average number of misclassified items*, i.e., the algorithm's average error count over different number of tests. To gain insight into the relative performance of both algorithms, we consider the function

$$\Delta(T) = \overline{M}_{\text{SCOMP}}(T) - \overline{M}_{\text{W-SCOMP}}(T),$$

where $\overline{M}_{\text{SCOMP}}(T)$ and $\overline{M}_{\text{W-SCOMP}}(T)$ denote the average number of misclassified items for SCOMP and W-SCOMP, respectively, at $T$ tests.

Figure 9 plots $\Delta(T)$ with an overlaid smoothing spline [41]. The difference is predominantly non-negative, indicating that SCOMP typically incurs higher misclassification error than W-SCOMP, thereby demonstrating the superior performance of W-SCOMP. The curves peak and then decline to zero since both algorithms converge when ample testing information is available. This behaviour is consistent with the algorithms' shared DD initialisation, as DD recovers more defectives with increasing $T$, the refinement stages have reduced impact.

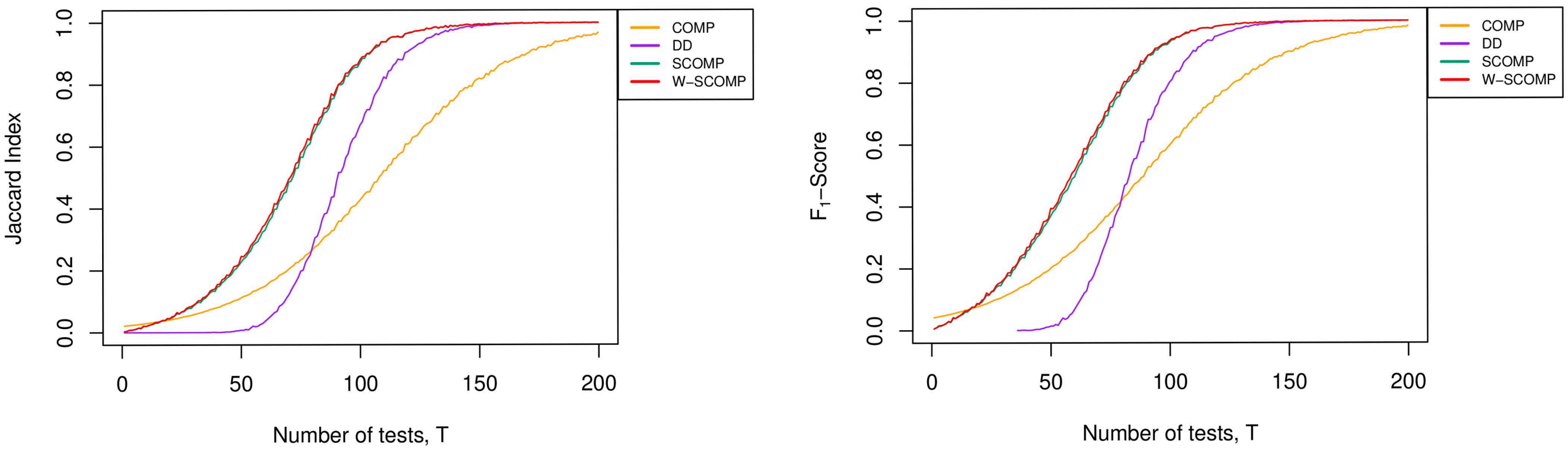

Fig. 8. Similarity metrics between true and estimated defective sets recovered for $T$ tests over $n = 1000$ independent simulations for each $T$ in a Bernoulli test design with $N = 500$ and $k = 10$: Jaccard index (left) and $F_1$-score (right).

Moreover, across different sparsity levels, the empirical results align with Theorem 3 and the observations in Figures 10 and 11, demonstrating that Weighted SCOMP more effectively differentiates between defective and non-defective items. As the problem becomes denser, this distinction becomes more pronounced and sustained over a broader range of tests.

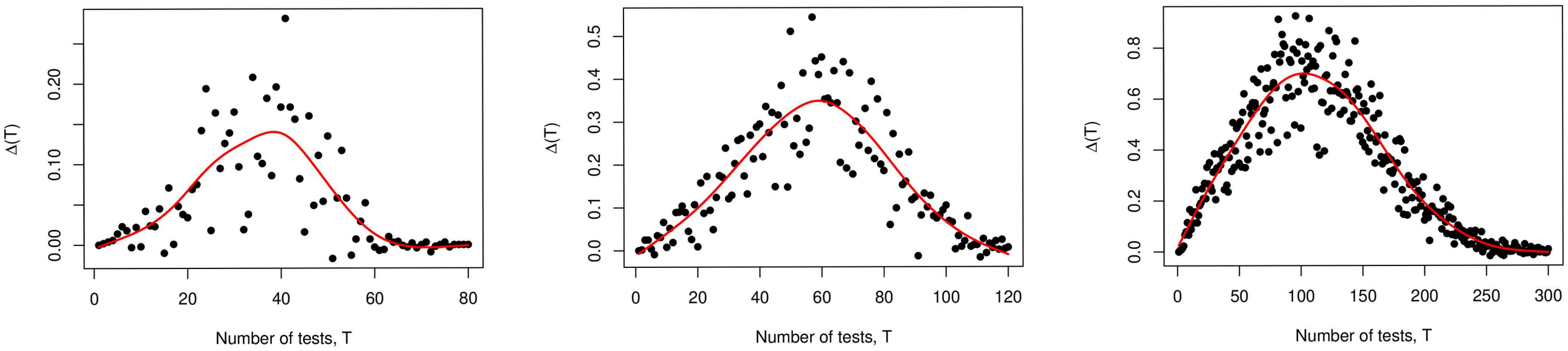

Fig. 9. Average misclassification difference with smoothing spline for $T$ tests over $n = 1000$ independent simulations for each $T$ in a Bernoulli test design with $N = 500$ and $k = 5$ (left), $k = 10$ (middle) and $k = 25$ (right).

## V. Conclusions

We studied the problem of noiseless non-adaptive group testing with the objective of maximising the success probability of exact defective-set recovery while minimising the number of tests. Motivated by contemporary applications (such as large-scale RT-PCR screening) we analysed classical decoders under several test designs and proposed a novel weighted decoding scheme.

Our main contributions are as follows:

1) We have proposed a new decoder, *Weighted SCOMP* (W-SCOMP), which incorporate test weights to better exploit test-design information.
2) We have introduced the signal-to-noise ratio (SNR) as a metric to compare the performance of noiseless non-adaptive group testing decoders.
3) We have given theoretical guarantees demonstrating that W-SCOMP strictly improves recovery performance relative to the classical decoders in the noiseless non-adaptive setting (see Theorem 3 and Section III-C).
4) We have empirically compared the performance of our proposed algorithms with the classical decoders across regimes of interest.
5) Extensive simulations have corroborated the theoretical results. W-SCOMP achieved the highest success probability among the considered algorithms, followed by SCOMP, DD and COMP (see Figures 1–9).

These findings advance the understanding of decoding methods for noiseless non-adaptive group testing, both in terms of practical performance and theoretical behaviour. They also suggest several concrete directions for future work. Immediate extensions include: (i) analysing the proposed weighted decoders under common noise models, (ii) deploying the methods on real experimental data and in application-specific pipelines.

# Appendix

## A. SNR

This appendix contains the full, explicit proofs of the main theorems and propositions that were given in the main text, as well as additional discussion and references to classical detection theory.

### 1) Full proof of Theorem 1:

**Theorem 1.** *Under Assumptions 1 and 2, the midpoint threshold classifier $\theta = \frac{1}{2}T(\mu_D + \mu_{ND})$ satisfies*

$$\mathbb{P}(\mathrm{err}_{\mathcal{T}}) \le \frac{4}{\mathrm{SNR}_{\mathcal{T}}^2}.$$

*Proof.* We bound the two error probabilities separately.

Under $H_D$ (i.e. the item is defective) the test statistic has mean $T\mu_D$. The false-negative probability (type II error) with threshold $\theta$ is

$$\begin{aligned}\mathbb{P}(\mathrm{FN}) &= \mathbb{P}(S_i^{(\mathcal{T})} < \theta \mid D) = \mathbb{P}(T\mu_D - S_i^{(\mathcal{T})} > T\mu_D - \theta \mid D)\\ &\le \mathbb{P}(T\mu_D - S_i^{(\mathcal{T})} > T\mu_D - \theta \mid D) + \mathbb{P}(T\mu_D - S_i^{(\mathcal{T})} < -(T\mu_D - \theta) \mid D)\\ &= \mathbb{P}(|S_i^{(\mathcal{T})} - T\mu_D| \le T\mu_D - \theta \mid D)\end{aligned}$$

By Chebyshev's inequality,

$$\mathbb{P}(|S_i^{(\mathcal{T})} - T\mu_D| \le T\mu_D - \theta \mid D) \le \frac{\mathrm{Var}[S_i^{(\mathcal{T})} \mid D]}{(T\mu_D - \theta)^2}.$$

Under Assumption 2, we have $\mathrm{Var}[S_i^{(\mathcal{T})} \mid D] = T\sigma_D^2$. Thus, for $\theta = \frac{1}{2}T(\mu_D + \mu_{ND})$, we have

$$\mathbb{P}(\mathrm{FN}) \le \frac{T\sigma_D^2}{T^2(\Delta\mu)^2/4} = \frac{4\sigma_D^2}{T(\Delta\mu)^2}.$$

Similarly, under $H_{ND}$ (non-defective) the false-positive probability (type I error) is

$$\mathbb{P}(\mathrm{FP}) = \mathbb{P}(S_i^{(\mathcal{T})} > \theta \mid ND) \le \mathbb{P}(|S_i^{(\mathcal{T})} - T\mu_{ND}| > \theta - T\mu_{ND} \mid ND) \le \frac{\mathrm{Var}[S_i^{(\mathcal{T})} \mid ND]}{(\theta - T\mu_{ND})^2} = \frac{4\sigma_{ND}^2}{T(\Delta\mu)^2}.$$

Adding the two error probabilities, the inequality of the statement is obtained

$$\mathbb{P}(\mathrm{err}_{\mathcal{T}}) \le \frac{4(\sigma_D^2 + \sigma_{ND}^2)}{T(\Delta\mu)^2} = \frac{4}{T \cdot \mathrm{SNR}_{per}^2}.$$

□

### 2) Full proof of Theorem 2:

**Theorem 2.** *Under Assumptions 1 and 2, the Bayes error,* $\mathrm{err}_{\mathcal{T}}^*$, *obeys*

$$\mathrm{err}_{\mathcal{T}}^* < \tfrac{1}{2}\exp\Big(-\frac{\mathrm{SNR}_{\mathcal{T}}^2}{4}\Big).$$

*Proof.* Let $f_1(x)$ and $f_2(x)$ be the density functions of $S_i^{(\mathcal{T})}$ under the conditions $c = D$ and $c = ND$ respectively. From Lemma 1,

$$(S_i^{(\mathcal{T})}|D) \sim \mathcal{N}(T\mu_D, T\sigma_D^2) \text{ and } (S_i^{(\mathcal{T})}|ND) \sim \mathcal{N}(T\mu_{ND}, T\sigma_{ND}^2).$$

Thus, the Bhattacharyya coefficient is

$$\rho = \int \sqrt{f_1(x)f_2(x)}\,dx = \exp(-B),$$

with

$$B = \frac{(T\mu_D - T\mu_{ND})^2}{4T(\sigma_D^2 + \sigma_{ND}^2)} + \frac{1}{4}\ln\left(\frac{T(\sigma_D^2 + \sigma_{ND}^2)}{2\sqrt{T^2\sigma_D^2\sigma_{ND}^2}}\right).$$

Simplifying,

$$\begin{aligned}B &= \frac{T\Delta\mu^2}{4(\sigma_D^2 + \sigma_{ND}^2)} + \frac{1}{4}\ln\left(\frac{\sigma_D^2 + \sigma_{ND}^2}{2\sqrt{\sigma_D^2\sigma_{ND}^2}}\right) = \frac{T \cdot \mathrm{SNR}_{per}^2}{4} + \frac{1}{4}\ln\left(\frac{\Sigma^2}{2\sqrt{\sigma_D^2\sigma_{ND}^2}}\right)\\ &\ge \frac{T \cdot \mathrm{SNR}_{per}^2}{4},\end{aligned}$$

since $\Sigma^2 \ge 2\sqrt{\sigma_D^2\sigma_{ND}^2}$.

Therefore, the Bayes error is bounded by $\mathrm{err}^*_{\mathcal{T}} \le \frac{1}{2}e^{-B} \le \frac{1}{2}\exp\left(-\frac{T\cdot \mathrm{SNR}^2_{per}}{4}\right)$, yielding the claimed expression. ☐

The proof of Theorem 2 invokes a Gaussian approximation for the class-conditional law of the aggregated score

$$S_i^{(\mathcal{T})} = \sum_{t\in\mathcal{T}} S_{t,i}, \qquad S_{t,i} = s(X_{t,i}, Y_t).$$

Hence, we require Lyapunov condition to be achieved in order to use Central Limit Theorem, this is proven in the following Lemma.

**Lemma 1.** *Under Assumptions 1 and 2, Lyapunov condition is achieved. Therefore, $(S_i^{(\mathcal{T})}|c)$ is asymptotically normal under each hypothesis $c \in \{D, ND\}$.*

*Proof.* For a fixed $c$. Since $s(X_{t,i}, Y_t)$ is upper-bounded, there exists $M < \infty$ such that $S_{t,i} \le M$ almost surely for all $t \in \mathcal{T}$. Therefore, denoting by $\sigma_T^2$ the variance of $S_i^{(\mathcal{T})}$, we have $|S_{t,i} - \mu_c| \le 2M$ and $\forall \delta > 0$

$$\frac{1}{\sigma_T^{2+\delta}}\mathbb{E}[|S_{t,i} - \mu_c|^{2+\delta}] \le \frac{(2M)^\delta}{\sigma_T^{2+\delta}}\sum_{t\in T}\mathrm{Var}(S_{t,i}) = \frac{(2M)^\delta}{\sigma_T^{2+\delta}}\cdot \sigma_T^2 = (2M)^\delta \sigma_T^{-\delta}.$$

Since $\sigma_T^2 \xrightarrow[T\to\infty]{} \infty$, then $(2M)^\delta\sigma_T^{-\delta} \xrightarrow[T\to\infty]{} 0$. Thus, Lyapunov condition is satisfied. ☐

*3) Full proof of Proposition 1 (DD non-locality): /*

**Proposition 1.** *DD's decision for an item depends on the global candidate set and thus cannot be represented as a function solely of per-test $(X_{t,i}, Y_t)$ contributions.*

*Proof.* DD first eliminates items that appear in any negative test. Among remaining candidates it looks for tests that are singletons relative to the candidate set: a positive test $t$ is a singleton for $i$ if $X_{t,i} = 1$ and $X_{t,j} = 0$ for every other candidate $j$. Whether $t$ is a singleton depends on which other items remained candidates after negative-test elimination, i.e. on the full set of negative-test outcomes across all items. Thus the predicate "$i$ is declared defective by DD" is a Boolean function of the entire design and outcomes, not a function only of $\{(X_{t,i}, Y_t)\}_t$. Consequently, no per-item additive scoring function depending only on $(X_{t,i}, Y_t)$ can represent DD in all cases. ☐

### *B. Derivations for the Weighted SCOMP moments*

*1) Defective item:* If item $i$ is defective then $Y_t = 1$ whenever $X_{t,i} = 1$, and (1) can be written as

$$W_{t,i}^{(D)} = \frac{1}{w_t}\mathbb{I}\{X_{t,i} = 1\} = \frac{1}{1+Z_t}\mathbb{I}\{X_{t,i} = 1\},$$

where $Z_t \sim \mathrm{Bin}(N-1, p)$ counts the other items included in the test. Hence

$$\mu_D^{(w)}(N,k) = \mathbb{E}\big[W_{t,i}^{(D)}\big] = \mathbb{P}(X_{t,i} = 1)\,\mathbb{E}\left[\frac{1}{1+Z_t}\right] = p\sum_{j=0}^{N-1}\frac{1}{1+j}\binom{N-1}{j}p^j(1-p)^{N-1-j}.$$

The sum can be simplified by the index shift $s = j+1$:

$$\begin{aligned}
\mu_D(N,k) = \mathbb{E}\left[\frac{1}{1+Z_t}\right] &= \sum_{j=0}^{N-1}\frac{1}{1+j}\binom{N-1}{j}p^j(1-p)^{N-1-j} \\
&= \frac{1}{Np}\sum_{j=0}^{N-1}\binom{N}{j+1}p^{j+1}(1-p)^{N-(j+1)} = \frac{1}{Np}\sum_{s=1}^{N}\binom{N}{s}p^s(1-p)^{N-s} \\
&= \frac{1-(1-p)^N}{Np},
\end{aligned} \tag{20}$$

therefore,

$$\mu_D^{(w)}(N,k) \;=\; p\,\mu_D(N,k) \;=\; \frac{1-(1-p)^N}{N}.$$

Similarly, the second moment is

$$\nu_D^{(w)}(N,k) = p\sum_{j=0}^{N-1}\frac{1}{(1+j)^2}\binom{N-1}{j}p^j(1-p)^{N-1-j},$$

which we leave in summation form for later numerical evaluation.

*2) Non-defective item:* If item $i$ is non-defective then among the remaining $N-1$ items there are $k$ defectives and $N-k-1$ non-defectives. We define

$$Z_{t,D} \sim \mathrm{Bin}(k,p) \text{ and } Z_{t,ND} \sim \mathrm{Bin}(N-k-1,p),$$

so that $Z_t = Z_{t,D} + Z_{t,ND}$ and $Z_{t,D}, Z_{t,ND}$ are independent. Let $A_t = \{Z_{t,D} \geq 1\}$ denote the event that test $t$ is positive; then

$$\mathbb{P}(A_t) = 1-(1-p)^k = q(k).$$

We have

$$W_{t,i}^{(ND)} = \frac{1}{1+Z_t}\mathbb{I}\{X_{t,i}=1\ ,\ A_t\},$$

so

$$\begin{aligned}\mu_{ND}^{(w)}(N,k) &= \mathbb{E}\big[W_{t,i}^{(ND)}\big] = \mathbb{P}(X_{t,i}=1,A_t)\,\mathbb{E}\left[\frac{1}{1+Z_t}\middle|X_{t,i}=1,A_t\right] \\ &= p\,q(k)\,\mu_{ND}(N,k),\end{aligned}$$

where

$$\mu_{ND}(N,k) = \mathbb{E}\left[\frac{1}{1+Z_t}\middle|Z_{t,D}\geq 1\right] = \frac{1}{q(k)}\sum_{h=1}^{k}\sum_{r=0}^{N-k-1}\frac{1}{1+h+r}\mathbb{P}(Z_{t,D}=h)\mathbb{P}(Z_{t,ND}=r).$$

Using the binomial probabilities yields the explicit double sum

$$\mu_{ND}(N,k) = \frac{1}{q(k)}\sum_{h=1}^{k}\sum_{r=0}^{N-k-1}\frac{1}{1+h+r}\binom{k}{h}\binom{N-k-1}{r}p^{h+r}(1-p)^{N-1-h-r} = \frac{1}{q(k)}\sum_{h=1}^{k}\sum_{r=0}^{N-k-1}\frac{1}{1+h+r}g(h,r), \tag{21}$$

where $g(h,r) = \binom{k}{h}\binom{N-k-1}{r}p^{h+r}(1-p)^{N-1-h-r}$.

Analogously, the second moment term $\nu_{ND}(N,k) = \mathbb{E}\big[(1+Z_t)^{-2} \mid Z_{t,D}\geq 1\big]$ leads to

$$\nu_{ND}(N,k) = \frac{1}{q(k)}\sum_{h=1}^{k}\sum_{r=0}^{N-k-1}\frac{1}{(1+h+r)^2}g(h,r), \tag{22}$$

and thus, $\nu_{ND}^{(w)}(N,k) = p\,q(k)\,\nu_{ND}(N,k)$.

### *C. Derivations for the unweighted SCOMP moments*

By definition,

$$U_{t,i} = \mathbb{I}\{X_{t,i}=1 \text{ and } Y_t = 1\}.$$

*1) Defective item:* If item $i$ is defective then $Y_t = 1$ whenever $X_{t,i}=1$, so (9) can be expressed as

$$U_{t,i}^{(D)} = \mathbb{I}\{X_{t,i}=1\}.$$

Therefore

$$\mu_D^{(u)}(k) = \mathbb{E}\big[U_{t,i}^{(D)}\big] = \mathbb{P}(X_{t,i}=1) = p,$$

and, since $U_{t,i}^{(D)}$ is an indicator,

$$\nu_D^{(u)}(k) = \mathbb{E}\big[(U_{t,i}^{(D)})^2\big] = p.$$

*2) Non-defective item:* If $i$ is non-defective, let $A_t = \{Y_t = 1\} = \{\text{at least one defective is included}\}$. Then

$$U_{t,i}^{(ND)} = \mathbb{I}\{X_{t,i}=1, A_t\},$$

and

$$\mu_{ND}^{(u)}(k) = \mathbb{E}\big[U_{t,i}^{(ND)}\big] = \mathbb{P}(X_{t,i}=1,A_t) = p\,q(k),$$

where $q(k) = 1-(1-p)^k$. Again, as $U_{t,i}^{(ND)}$ is an indicator,

$$\nu_{ND}^{(u)}(k) = \mathbb{E}\big[(U_{t,i}^{(ND)})^2\big] = p\,q(k).$$

### D. Full proof of Theorem 3

For convenience, we consider the normalized per-test SNR ratios

$$R_W(N,k) = \frac{\mathbb{E}[W_D] - q(k)\mathbb{E}[W_{ND}]}{\sqrt{\mathbb{E}[W_D^2] + q(k)\mathbb{E}[W_{ND}^2] - (1+q(k))(\mathbb{E}[W_D]^2 + q(k)\mathbb{E}[W_{ND}]^2)}} \tag{23}$$

and

$$R_U(k) = \frac{1-q(k)}{\sqrt{1-p+q(k)(1-pq(k))}}, \tag{24}$$

where we use the simplified notation $W_D = (1+Z_t)^{-1}$ and $W_{ND} = (1+Z_{t,D}+Z_{t,ND})^{-1} \mid Z_{t,D} \geq 1)$. So that the corresponding per-test signal-to-noise ratios are

$$\mathrm{SNR}_W(N,k) = \frac{p}{\sqrt{p}} R_W(N,k) \text{ and } \mathrm{SNR}_U(k) = \frac{p}{\sqrt{p}} R_U(k).$$

*1) Moments of $W_D$ and $W_{ND}$:* Taking into account that $p = 1/(k+1)$ and $q(k) = 1-(1-p)^k$, and using the generalized Vandermonde's identity

$$\binom{N-1}{j} = \sum_{\substack{h\geq 0,\ r\geq 0\\ h+r=j}} \binom{k}{h}\binom{N-k-1}{r},$$

one obtains following expressions of the first and second moments:

$$\mathbb{E}[W_D] = \mu_D(N,k) = \frac{k+1}{N}\Big(1-\Big(\frac{k}{k+1}\Big)^N\Big) = \sum_{h=0}^{k}\sum_{r=0}^{N-k-1} \frac{1}{1+h+r} g(h,r), \tag{25}$$

$$\mathbb{E}[W_D^2] = \nu_D(N,k) = \sum_{j=0}^{N-1} \frac{1}{(1+j)^2}\binom{N-1}{j} p^j(1-p)^{N-1-j} = \sum_{h=0}^{k}\sum_{r=0}^{N-k-1} \frac{1}{(1+h+r)^2} g(h,r), \tag{26}$$

$$\mathbb{E}[W_{ND}] = \mu_{ND}(N,k) = \frac{1}{q(k)}\sum_{h=1}^{k}\sum_{r=0}^{N-k-1} \frac{1}{1+h+r} g(h,r), \tag{27}$$

$$\mathbb{E}[W_{ND}^2] = \nu_{ND}(N,k) = \frac{1}{q(k)}\sum_{h=1}^{k}\sum_{r=0}^{N-k-1} \frac{1}{(1+h+r)^2} g(h,r). \tag{28}$$

*2) Evaluation of the numerator of* (23)*:* From (25) and (27), the numerator of (23) simplifies by cancelling all terms with $h \geq 1$:

$$\begin{aligned}\mathbb{E}[W_D] - q(k)\mathbb{E}[W_{ND}] &= \sum_{h=0}^{k}\sum_{r=0}^{N-k-1} \frac{1}{1+h+r} g(h,r) - \sum_{h=1}^{k}\sum_{r=0}^{N-k-1} \frac{1}{1+h+r} g(h,r) \\ &= \sum_{r=0}^{N-k-1} \frac{1}{1+r}\binom{N-k-1}{r} p^r(1-p)^{N-1-r}.\end{aligned}$$

Re-indexing (set $s = r+1$) and simplifying, it leads to the closed form

$$\mathbb{E}[W_D] - q(k)\mathbb{E}[W_{ND}] = \frac{(1-p)^k}{(N-k)p}\Big(1-(1-p)^{N-k}\Big), \tag{29}$$

which is strictly positive for all integers $0 < k < N$.

*3) Explicit expression for $\mathbb{E}[W_{ND}]$:* Solving (29) for $\mathbb{E}[W_{ND}]$ using the closed form $\mathbb{E}[W_D] = \frac{1-(1-p)^N}{Np}$ we have from (20), produces the explicit formula:

$$\mathbb{E}[W_{ND}] = \frac{k+1}{1-(k/(k+1))^k}\left(\frac{1-(k/(k+1))^N}{N} - \frac{(k/(k+1))^k\big(1-(k/(k+1))^{N-k}\big)}{N-k}\right). \tag{30}$$

*4) A Jensen lower bound for second moments:* The convexity of $g(x) = (1+x)^{-2}$ gives a simple Jensen lower bound for $\mathbb{E}[W_D^2] = \mathbb{E}[g(Z_t)]$:

$$\mathbb{E}[W_D^2] \geq g(\mathbb{E}[Z_t]) = \frac{1}{(1+(N-1)p)^2} = \left(\frac{k+1}{N+k}\right)^2.$$

Similarly, for $W_{ND}$,

$$\mathbb{E}[W_{ND}^2] \geq \left(\frac{(k+1)\big(1-\frac{k}{k+1}\big)^k}{k+N\big(1-\frac{k}{k+1}\big)^k}\right)^2.$$

These bounds are used to control the denominator of $R_W$.

*5) Second-moment sum:* The sum $\mathbb{E}[W_D^2] + q(k)\mathbb{E}[W_{ND}^2] = 2\mathbb{E}[W_D^2] - (\mathbb{E}[W_D^2] - q(k)\mathbb{E}[W_{ND}^2])$, and taking into account (26) and (28)

$$\begin{aligned}
\mathbb{E}[W_D^2] + q(k)\mathbb{E}[W_{ND}^2] &= 2\sum_{j=0}^{N-1}\binom{N-1}{j}\frac{1}{j+1}p^j(1-p)^{N-j-1} \\
&- \left(\sum_{h=0}^{k}\sum_{r=0}^{N-k-1}\frac{1}{(1+h+r)^2}g(h,r) - \sum_{h=1}^{k}\sum_{r=0}^{N-k-1}\frac{1}{(1+h+r)^2}g(h,r)\right) \\
&= \frac{2}{Np}\sum_{s=1}^{N}\binom{N}{s}\frac{1}{s}p^s(1-p)^{N-s} - \sum_{r=0}^{N-k-1}\frac{1}{(1+r)^2}\binom{N-k-1}{r}p^r(1-p)^{N-1-r} \\
&= \frac{2}{Np}\sum_{s=1}^{N}\binom{N}{s}\frac{1}{s}p^s(1-p)^{N-s} - \frac{1}{p(N-k)}\sum_{s=1}^{N-k}\frac{1}{s}\binom{N-k}{s}p^s(1-p)^{N-s},
\end{aligned} \tag{31}$$

which is used in Appendix D6 below.

*6) Final form:* Squaring (23) and (24), the inequality (17), $R_W(N,k) \geq R_U(k)$,is equivalent to inequality (18). After rearrangement and multiplication by $(1-q(k))^2$, we obtain the form (19) where the functions $f_j(k)$ are

$$\begin{aligned}
f_1(k) &= 1 - 2p\,q(k) + q(k), \\
f_2(k) &= -2q(k)\left(1 - p + q(k)\left(1 - pq(k)\right)\right), \\
f_3(k) &= q^2(k)\left(1 - 2p\,q(k) + q(k)\right), \\
f_4(k) &= -(1-q(k))^2.
\end{aligned} \tag{32}$$

Substituting the moment sums (25)–(28) and the expressions (29)–(31) into (19), one can obtain the following expanded function

$$\begin{aligned}
f(N,k) = \frac{k+1}{N^2}&\left(2k - (k-1)\left(\frac{k}{k+1}\right)^k\right)\left(1 - \left(\frac{k}{k+1}\right)^N\right)^2 \\
&- \frac{k+1}{N}\frac{1-\left(\frac{k}{k+1}\right)^N}{1-\left(\frac{k}{k+1}\right)^k}\left(\frac{1-\left(\frac{k}{k+1}\right)^N}{N} - \frac{\left(\frac{k}{k+1}\right)^k(1-\left(\frac{k}{k+1}\right)^{N-k})}{N-k}\right) \\
&\times\left(4k - (6k-2)\left(\frac{k}{k+1}\right)^k + (2k-4)\left(\frac{k}{k+1}\right)^{2k} + 2\left(\frac{k}{k+1}\right)^{3k}\right) \\
&+ \frac{k+1}{\left(1-\left(\frac{k}{k+1}\right)^k\right)^2}\left(\frac{1-\left(\frac{k}{k+1}\right)^N}{N} - \frac{\left(\frac{k}{k+1}\right)^k(1-\left(\frac{k}{k+1}\right)^{N-k})}{N-k}\right)^2 \\
&\times\left(2k - (5k-1)\left(\frac{k}{k+1}\right)^k + (4k-2)\left(\frac{k}{k+1}\right)^{2k} - (k-1)\left(\frac{k}{k+1}\right)^{3k}\right) \\
&- (k+1)\left(\frac{k}{k+1}\right)^{N+2k}\left(\frac{2}{N}\sum_{s=1}^{N}\binom{N}{s}\frac{1}{s\,k^s} - \frac{1}{N-k}\sum_{s=1}^{N-k}\binom{N-k}{s}\frac{1}{s\,k^s}\right),
\end{aligned}$$

and hence, (17) is equivalent to $f(N,k) \geq 0$.

*7) Sign of $f(N,k)$:* The functions given in (32) satisfy simple sign and monotonicity properties for $k \geq 1$:

- $f_1(k) = 1 - 2pq(k) + q(k) > 0$ and is increasing in $k$.
- $f_2(k) = -2q(k)\big(1 - p + q(k)(1 - pq(k))\big) < 0$ and is strictly decreasing in $k$, numerically it takes values in the interval $\big(-4 + 6e^{-1} - 2e^{-2}, -7/8\big]$.
- $f_3(k) = q^2(k)\big(1 + q(k) - 2pq^2(k)\big) > 0$ and is increasing in $k$.
- $f_4(k) = -(1-q(k))^2 < 0$ and is strictly increasing in $k$ with values in $\big[-1/4, -e^{-2}\big)$.

From (25) and (30) we have $\lim_{N\to\infty}\mathbb{E}[W_D] = 0$ and $\lim_{N\to\infty}\mathbb{E}[W_{ND}] = 0$. In addition, expression (31) also tends to 0 as $N \to \infty$. Hence, $\lim_{N\to\infty} f(N,k) = 0$ for each fixed $k \geq 1$. Together with the sign and monotonicity properties recorded earlier, this limit behaviour is used to exclude the possibility of $f(N,k)$ taking negative values on the admissible domain.

Although the preceding chain of exact reductions and bounds leaves $f(N,k)$ in a form that resists a short closed-form lower bound, the exact expression is explicit and readily evaluated. Figures 10 and 11 plot $f(N,k)$ for ranges of $(N,k)$, empirically illustrating that $f(N,k) > 0$ throughout the admissible region.

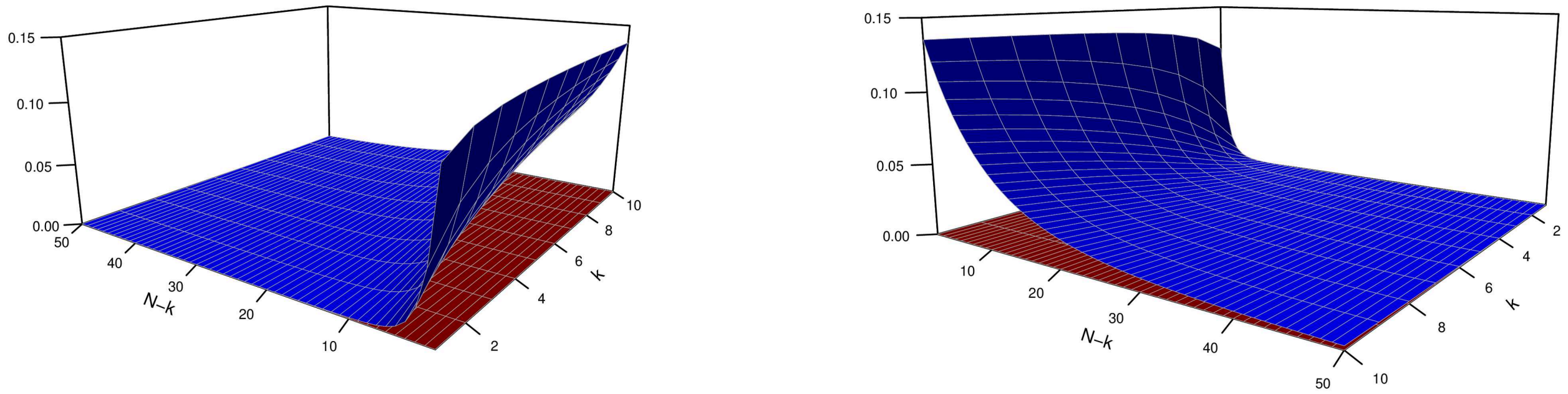

Fig. 10. Plots of $f(N,k)$ for $k=1,...,10$ and $N=k+1,...,k+50$, and the red surface $z(x,y)=0$.

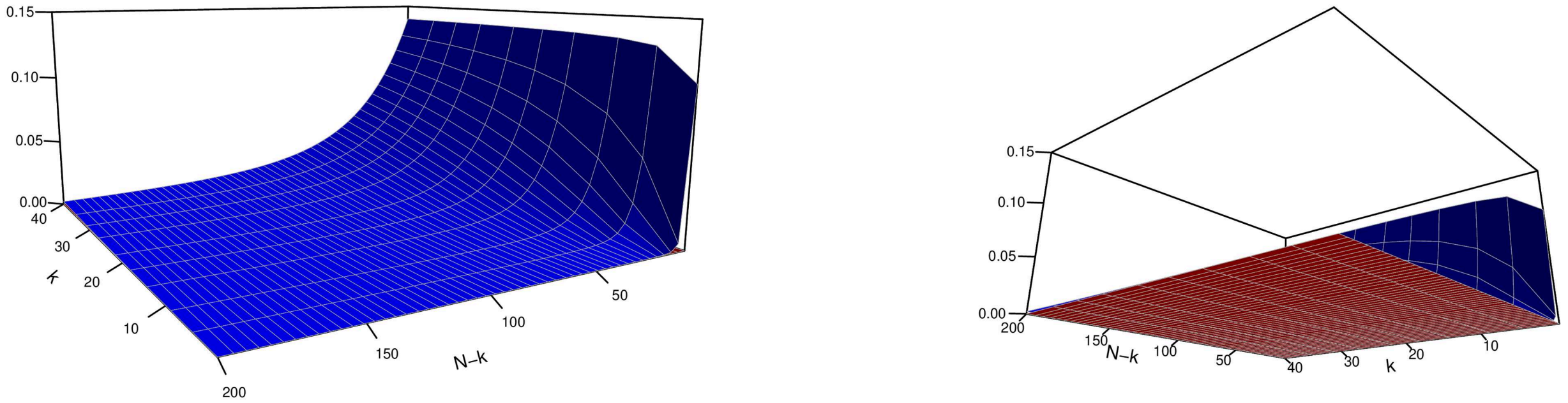

Fig. 11. Plots of $f(N,k)$ for $k=1,...,40$ and $N=k+1,...,k+200$, and the red surface $z(x,y)=0$.

## *E. Proof of Corollary 1*

Corollary 1 indicates that the aggregated weighted score concentrates at least as sharply as the aggregated unweighted score.

*Proof.* We remark that the aggregated scores $W_i^{(\mathcal{T})}=\sum_{t\in\mathcal{T}} W_{t,i}$ and $U_i^{(\mathcal{T})}=\sum_{t\in\mathcal{T}} U_{t,i}$ are sums of independent random variables, with $\mathbb{E}[W_{t,i}]=\mu^{(w)}(N,k)$ and $\mathbb{E}[U_{t,i}]=\mu^{(u)}(k)$, and variances $\mathrm{Var}(W_{t,i})=\sigma_w^2(N,k)$ and $\mathrm{Var}(U_{t,i})=\sigma_u^2(k)$.

The maximum deviation for each weighted test, $M_w$, is given by

$$M_w=\max_t \left|W_{t,i}-\mu^{(w)}(N,k)\right|,$$

and similarly for the unweighted score,

$$M_u=\max_t \left|U_{t,i}-\mu^{(u)}(k)\right|.$$

In our group testing model, note that $W_{t,i}$ is of the form $1/w_t$ (when the test $t$ is positive and item $i$ is included) and thus it is upper bounded by 1, while $U_{t,i}$ is the indicator function. In practice, due to the averaging effect of the weight $1/w_t$ (with $w_t\geq 1$), one has $M_w\leq M_u$.

Then, applying Bernstein's inequality [42], for any $\varepsilon>0$, we have

$$\mathbb{P}\left(\left|W_i^{(\mathcal{T})}-T\,\mu^{(w)}(N,k)\right|\geq\varepsilon\right)\leq 2\exp\left(-\frac{\varepsilon^2}{2T\,\sigma_w^2(N,k)+\frac{2M_w\,\varepsilon}{3}}\right), \tag{33}$$

and

$$\mathbb{P}\left(\left|U_i^{(\mathcal{T})}-T\,\mu^{(u)}(k)\right|\geq\varepsilon\right)\leq 2\exp\left(-\frac{\varepsilon^2}{2T\,\sigma_u^2(k)+\frac{2M_u\,\varepsilon}{3}}\right). \tag{34}$$

Hence, the exponent in (33) is larger than that in (34). In turn, the weighted score $W_i^{(\mathcal{T})}$ exhibits a sharper concentration around its mean compared to $U_i^{(\mathcal{T})}$.

Furthermore, Theorem 3 implies that

$$\mathrm{SNR}_W(N,k)=\frac{\Delta\mu^{(w)}(N,k)}{\sqrt{\sigma_w^2(N,k)}}\geq\frac{\Delta\mu^{(u)}(k)}{\sqrt{\sigma_u^2(k)}}=\mathrm{SNR}_U(k),$$

so that, for fixed means, we have $\sigma_w^2\leq\sigma_u^2$, and the bounded deviation $M_w$ is no larger than $M_u$. Thus, the denominator in the exponential of (33) is smaller than (or equal to) that in (34). This implies that for any deviation $\varepsilon$, the tail probability bound for the weighted score decays at a faster exponential rate than that for the unweighted score. □

## Acknowledgments

Part of this work was developed for the award of the Master of Engineering in Mathematics and Computer Science at the University of Bristol, supervised by Professor Oliver Johnson. I thank him for his advice and support, and the revision of this manuscript.